\documentstyle[prc,aps,epsf,preprint,eqsecnum]{revtex}

\tighten

\begin{document}
\newcommand{\hgo}{$^{190}$Hg}
\newcommand{\hgt}{$^{192}$Hg}
\newcommand{\hgf}{$^{194}$Hg}
\newcommand{\J}{{\cal J}^{(2)}}
\newcommand{\JJ}{{\cal J}^{(1)}}
\newcommand{\w}{\omega_{\rm rot}}
\newcommand{\bra}[1]{\langle #1 |}
\newcommand{\ket}[1]{| #1 \rangle}

\draft
\preprint{TASCC-P-95-37}

\title{Microscopic Structure of High-Spin Vibrational Excitations\\
in Superdeformed $^{190,192,194}$Hg}
\author{Takashi~Nakatsukasa\footnote{
E-mail : {\tt nakatsuk@cu1.crl.aecl.ca}}}
\address{AECL, Chalk River Laboratories, Chalk River,
Ontario K0J 1J0, Canada}
\author{Kenichi~Matsuyanagi}
\address{Department of Physics, Kyoto University, Kyoto 606--01, Japan}
\author{Shoujirou~Mizutori}
\address{
Department of Mathematical Physics, Lund Institute of Technology,
Box 118, S-22100, Lund, Sweden}
\author{Yoshifumi~R.~Shimizu}
\address{
Department of Physics, Kyushu University, Fukuoka 812, Japan}

\maketitle

\bigskip
\begin{abstract}
Microscopic RPA calculations based on the cranked shell model are
performed to investigate the quadrupole and octupole correlations
for excited superdeformed bands in \hgo, \hgt, and \hgf.
The $K=2$ octupole vibrations are predicted to be the lowest excitation
modes at zero rotational frequency.
At finite frequency, however,
the interplay between rotation and vibrations
produces different effects depending on
neutron number:
The lowest octupole phonon is rotationally aligned in \hgo,
is crossed by the aligned two-quasiparticle bands in \hgt,
and retains the $K=2$ octupole vibrational character
up to the highest frequency in \hgf.
The $\gamma$ vibrations are predicted to be higher in energy
and less collective than the octupole vibrations.
{}From a comparison with the experimental
dynamic moments of inertia,
a new interpretation of the observed excited bands invoking the $K=2$
octupole vibrations is proposed,
which suggests those octupole vibrations may be prevalent in SD Hg nuclei.
\end{abstract}

%\pacs{PACS number(s): 21.10.Re, 21.60.Jz, 27.80.+w}
%\bigskip
\newpage
%\narrowtext

\section{Introduction}
\label{sec: intro}

Theoretical and experimental studies of collective vibrational states
built on the superdeformed (SD) yrast band are
open topics of interest
in the field of high-spin nuclear structure.
Since the large deformation and rapid rotation of SD bands may
produce a novel shell structure,
we expect that surface vibrations will exhibit quite different features
from those found in spherical and normal-deformed nuclei.
According to our previous work\cite{MSM90,NMM92,NMM93,Miz93,Nak95},
low-lying octupole vibrations are more important than quadrupole vibrations
when the nuclear shape is superdeformed.
Strong octupole correlations in SD states have been also suggested
theoretically in Refs.\cite{DWS90,Abe90,HA90,Bon91,LDR91,ND92,Ska92,Ska93}.
Experimentally, octupole correlations in SD states have been suggested
for $^{152}$Dy\cite{Dag95}, $^{193}$Hg\cite{Cul90}
and $^{190}$Hg\cite{Cro94,Cro95}.
We have reported theoretical calculations
corresponding to these data
for $^{193}$Hg\cite{NMM93} and $^{152}$Dy\cite{Nak95}.
In this paper, we discuss the quadrupole and octupole correlations
for \hgo\ (which have been
partially reported in Refs.\cite{Cro95,Wil95,Nak95_2})
and for the neighboring SD nuclei $^{192,194}$Hg.

We have predicted the low-lying $K=2$ octupole vibrations
for SD Hg isotopes $^{190,192,194}$Hg
($E_x\sim 1$MeV)\cite{NMM93,Miz93}.
These predictions differ from the results of
generator-coordinate-method (GCM) calculations\cite{Ska93}
in which the $K=0$ octupole state is predicted to
be the lowest in SD \hgt\ and the excitation energies are
significantly higher ($E_x\sim 2$MeV) than in our predictions.
Experimentally \cite{Cro95},
the routhians of the lowest octupole state
decrease with the rotational frequency, for example from
$E'_x\approx 0.7$MeV to 0.3MeV as
$\hbar\w$ goes from 0.25 to 0.35MeV, therefore
to compare the theoretical routhians directly with the experimental
ones, we need to calculate them at finite rotational frequency.
For this purpose, the cranked shell model extended by the random-phase
approximation (RPA) provides us with
a powerful tool to investigate collective excitations at high angular
momentum.

A great advantage of this model is its ability to take into account
effects of the Coriolis coupling on the collective vibrational motions
in a rapidly rotating system.
Since in the normal-deformed nuclei it is known
that Coriolis coupling effects are
important even for the $3^-$ octupole states\cite{NV70},
one may expect strong Coriolis mixing for high-spin octupole
states built on the SD yrast band.
On the other hand, our previous calculations suggested
weak Coriolis mixing for the lowest octupole state
in \hgt\cite{NMM93} and $^{152}$Dy\cite{Nak95}.
This may be because
the angular momentum of the octupole phonon is strongly coupled
to the symmetry axis due to
the large deformation of the SD shape.
Generally speaking, Coriolis mixing is expected to occur more easily
in nuclei with smaller deformation.
However this expectation may not hold
for octupole bands in all SD nuclei
because Coriolis mixing
depends on the shell structure.  In this paper
we find a significant difference in the Coriolis
mixing between an octupole band in \hgo\ and the other bands.

Another advantage of this model is that it gives us a unified
microscopic description of
collective states, weakly-collective states, and
non-collective two-quasiparticle excitations.
A transition of the octupole vibrations into aligned two-quasiparticle
bands at
high-spin in normal-deformed nuclei
has been predicted by Vogel\cite{Vog76}.
In Ref.\cite{Nak95_2}, this transition is discussed in the context of
experimental data on rare-earth and actinide nuclei,
and a damping of octupole collectivity at high spin was suggested.
Since similar phenomena may happen to octupole vibrations in SD states,
it is important that our model describes the interplay
between collective and non-collective excitations.

Recent experimental studies reveal a number of interesting features
of excited SD bands in even-even Hg isotopes.
In \hgo, almost constant dynamic moments of inertia $\J$ have been
observed by Crowell et al.\cite{Cro94}.
Ref.\cite{Cro95} has established the relative excitation energy
of this band and confirmed the dipole character of
the decay transitions into the yrast SD band.
This band has been interpreted as an
octupole vibrational band.
Two more excited bands in \hgo\ have been observed recently
by Wilson et al.\cite{Wil95},
one of which shows a sharp rise of $\J$ at low frequency.
In \hgt, Fallon et al.\cite{Fal95} have reported two excited bands which
exhibit peaks in $\J$ at high frequency.
In contrast with these atypical $\J$ behaviors, two excited bands in
\hgf\ originally
observed by Riley et al.\cite{Ril90} and extended
by Cederwall et al.\cite{Ced95}
show a smooth increase with rotational frequency.
We show in this paper that this $\J$ behavior
can be explained with a single theoretical model
which microscopically takes into account
shape vibrations and the Coriolis force.

The purpose of this paper is to
present the RPA method based on the cranked shell model
and its ability to describe a variety of nuclear properties including
shape vibration at large deformation and high spin.
We propose a plausible interpretation for the microscopic structure of
excited SD bands in $^{190,192,194}$Hg,
and show that octupole bands may be more prevalent than expected in these
SD nuclei.
Section \ref{sec: model} presents a description of the model,
in which
we stress our improvements to the cranked Nilsson potential
and to the coupled RPA method in a rotating system.
Section \ref{sec: detail} presents details of the calculation
in which the pairing and effective interactions are discussed.
The results for the excited SD \hgo, \hgt, and \hgf\
are presented in section \ref{sec: results}, and compared with the
experimental data in section \ref{sec: comparison}.
The conclusions are summarized in section \ref{sec: conclusion}.

%%%%%%%%%%%%%%%% Model description %%%%%%%%%%%%%%%%%%%%%%%%%%%%%%%%%%%%
\section{Theoretical framework}
\label{sec: model}

The theory of the cranked shell model extended by the random-phase
approximation (RPA) was first developed
by Marshalek\cite{Mar75} and has been applied to high-spin
$\beta$ and  $\gamma$ vibrational bands\cite{EMR80,SM83,SM95} and
to octupole bands\cite{RER86,MSM90,NMM92,NMM93,Miz93,Nak95}.
Since this theory is suitable for describing the collective
vibrations built on deformed high-spin states,
it is very useful for investigating vibrational
motion built on the SD yrast band.

%%%%%%%%%%%%%%%% Cranked Nilsson %%%%%%%%%%%%%%%%%%%%%%%%%%%%%%%%%%%%
\subsection{The cranked Nilsson potential with the local Galilean
invariance}

We start with a rotating mean field with a rotational frequency
$\w$ described by
\begin{equation}
\label{sp-potential}
h_{\rm s.p.}=h_{\rm Nilsson}
+ \Gamma_{\rm pair} - \w J_x \ + h_{\rm add},
\end{equation}
where $h_{\rm Nilsson}$ is a standard Nilsson potential defined in
single-stretched coordinates, $r'_i=\sqrt{\frac{\omega_i}{\omega_0}} r_i$
and $p'_i=\sqrt{\frac{\omega_0}{\omega_i}} p_i$ ($i=x,y,z$),
\begin{equation}
\label{Nilsson}
h_{\rm Nilsson}=\left( \frac{\omega_i}{\omega_0} \right)
\left( \frac{{{\bf p}'}^2}{2M}
+\frac{M\omega_0^2}{2} {{\bf r}'}^2 \right)
+v_{ll}\left({{\bf l}'}^2
 -\langle {{\bf l}'}^2 \rangle_N\right)
+v_{ls}{\bf l}'\cdot {\bf s} \ ,
\end{equation}
where ${\bf l}'={\bf r}'\times {\bf p}'$.
The pairing field $\Gamma_{\rm pair}$ is defined by
\begin{equation}
\label{pair-field}
\Gamma_{\rm pair} = -\sum_{\tau=n,p}\Delta_\tau
\left(P_\tau^\dagger + P_\tau\right) -\sum_{\tau=n,p} \lambda_\tau N_\tau
\ ,
\end{equation}
where $P_\tau = \sum_{k\in\tau,k>0} c_{\bar{k}} c_k$
and $N_\tau=\sum_{k\in\tau} c_k^\dagger c_k$ are the monopole-pairing and
number operators, respectively.
In section~\ref{sec: detail-MF}, we discuss the details of the pairing
field used in the calculations.

A standard cranked Nilsson potential has the disadvantage
that it overestimates the
moments of inertia compared to a Woods-Saxon potential.
This problem comes from the spurious velocity-dependence associated with
the ${\bf l^2}$-term in the Nilsson potential
which is absent for Woods-Saxon.
If the mean-field potential is velocity independent,
the local velocity distribution in the rotating nucleus remains isotropic
in velocity space, which means that the flow pattern becomes the same as
for a rigid-body rotation\cite{BM75}.
However, in the cranked Nilsson potential,
this isotropy of the velocity distribution is significantly broken due to
the ${\bf l^2}$-term.
Thus the Coriolis force introduces a spurious flow
in the rotating coordinate system,
proportional to the rotational frequency.
This spurious effect can be compensated by
an additional term that
restores the local Galilean invariance.
This additional term is obtained by substituting
(the local Galilean transformation)
\begin{equation}
{\bf p} \longrightarrow
{\bf p}-M\left(\mbox{\boldmath $\w$}\times {\bf r}\right)\ ,
\end{equation}
in the ${\bf ls}$- and ${\bf l^2}$-terms of the Nilsson potential.
This prescription was suggested by Bohr and Mottelson\cite{BM75}, and
developed by Kinouchi and Kishimoto\cite{Kin88}.
For a momentum-dependent potential $V({\bf r},{\bf p})$,
\begin{eqnarray}
V({\bf r},{\bf p}) + h_{\rm add} &=&
V\left({\bf r},{\bf p}
-M\left(\mbox{\boldmath $\w$}\times {\bf r}\right)\right) \ ,\\
&\approx& V({\bf r},{\bf p}) - \w M \left( y \frac{\partial}{\partial p_z}
    - z \frac{\partial}{\partial p_y}\right) V({\bf r},{\bf p})\ ,\\
 &=& V({\bf r},{\bf p}) + \frac{i}{\hbar} \w M \left( y \left[ z,V\right]
    - z \left[ y,V\right] \right) \ ,
\end{eqnarray}
where we assume uniform rotation around the $x$-axis,
$\mbox{\boldmath $\w$}=(\w,0,0)$.
Following this prescription,
the additional term $h_{\rm add}$ in eq.(\ref{sp-potential})
is obtained for the Nilsson potential (\ref{Nilsson}),
\begin{equation}
\label{additional-term}
h_{\rm add}=-\frac{\w}{\sqrt{\omega_y\omega_z}}\left\{
v_{ll} \left(2M\omega_0 {{\bf r}'}^2
 - \hbar\left(N_{\rm osc}+\frac{3}{2} \right)\right) l'_x
+v_{ls}M\omega_0\left({{\bf r}'}^2 s_x
 - r'_x \left({\bf r}'\cdot {\bf s}\right)\right)\right\} \ .
\end{equation}
Note that the term proportional to
$\left( N_{\rm osc}+\frac{3}{2} \right)$
in eq.(\ref{additional-term})
comes from the velocity-dependence of $\langle {{\bf l}'}^2 \rangle_N$
in eq.(\ref{Nilsson}).  This result,
eq.(\ref{additional-term}), has been applied to the SD bands in
$^{152}$Dy
\cite{Nak95} where the
single-particle routhians were found to be
very similar to those obtained
by using the Woods-Saxon potential.
In Fig.~\ref{dy152_J}, moments of inertia for SD $^{152}$Dy
calculated with and without
the additional term (\ref{additional-term}) are displayed.
Since the effects of the mixing among the major oscillator shells
$N_{\rm osc}$ are neglected in calculating our routhians,
kinematic ($\JJ$) and dynamic ($\J$) moments of inertia are obtained
by adding the contributions of the $N_{\rm osc}$-mixing effects
to the values calculated without them:
\begin{eqnarray}
\JJ &=&
 \frac{\langle J_x \rangle}{\w} + {\it\Delta}{\cal J}_{\rm Inglis} \ ,\\
\J &=&
 \frac{d\langle J_x \rangle}{d\w} + {\it\Delta}{\cal J}_{\rm Inglis} \ ,\\
{\it\Delta}{\cal J}_{\rm Inglis} &=&
 {\cal J}_{\rm Inglis}-{\cal J}_{\rm Inglis}^{{\it\Delta}N=0}
 = 2 \sum_{n({\it\Delta}N=2)}
   \frac{|\bra{n}J_x \ket{0}|^2}{E_n-E_0} \ ,
\end{eqnarray}
where ${\it\Delta}{\cal J}_{\rm Inglis}$ is difference between
the Inglis moments of inertia with and without the ${\it\Delta}N_{\rm osc}=2$
contributions\cite{SVB90}.
The $\JJ$ and $\J$ values calculated with the additional term
are very close to
the rigid-body value at low frequency, which means that the spurious effects
of the ${\bf l}^2$-term have been removed.
Note that the abscissa of Fig.~\ref{dy152_J} corresponds to the ``bare''
rotational frequency without renormalization.
The drastic reduction of $\JJ$ and $\J$ at high frequency
is corrected by the additional term, and this is seen to be
important in reproducing the experimental $\J$-behavior of the
yrast SD band.

%%%%%%%%%%%%%%%% coupled RPA %%%%%%%%%%%%%%%%%%%%%%%%%%%%%%%%%%%%
\subsection{The RPA in the rotating frame}

The residual interactions are assumed to be in a separable form,
\begin{equation}
H_{\rm int} = -\frac{1}{2} \sum_{\rho,\alpha}
              \chi_\rho R_\rho^\alpha R_\rho^\alpha \ ,
\label{residual-int}
\end{equation}
where $R_\rho^\alpha$ are one-body Hermitian operators, and $\chi_\rho$
are coupling strengths.
The indices $\alpha$ indicate the signature quantum numbers
($\alpha=0,1$) and $\rho$ specifies other modes.
In this paper, we take as $R_\rho^\alpha$ the monopole pairing and
the quadrupole operators for positive-parity states, and
the octupole and the isovector dipole operators for negative-parity
states (see eq.(\ref{operator_R})).
Since the $K$-quantum number is not conserved at finite rotational
frequency, it is more convenient to make the multipole operators
have good signature quantum numbers.
In general, the Hermitian multipole (spin-independent) operators
with good signature quantum numbers are constructed by
\begin{equation}
Q_{\lambda K}^\alpha =
 \frac{i^{\lambda+\alpha+K}}{\sqrt{2(1+\delta_{K0})}}
 \left( r^\lambda Y_{\lambda K}
 + (-)^{\lambda+\alpha} r^\lambda Y_{\lambda -K} \right) \quad\quad
(K\ge 0)
\label{multipole-op}
\end{equation}
where the spherical-harmonic functions $Y_{\lambda K}$
are defined with respect to the symmetry ($z$-) axis.
All multipole operators are defined in
doubly-stretched coordinates,
($r^{''}_i=\frac{\omega_i}{\omega_0} r_i$),
which can be regarded as an improved version of the conventional
multipole interaction.
Sakamoto and Kishimoto\cite{SK89} have shown that
at the limit of the harmonic-oscillator potential (at $\w=0$),
it guarantees nuclear self-consistency\cite{BM75},
restoration of the symmetry broken in the mean field,
and separation of the spurious solutions.
The coupling strengths $\chi_\rho$ should be determined by
the self-consistency condition between the density distribution and
the single-particle potential (see section \ref{sec: detail-RPA}
for details).

To describe vibrational excitations in the RPA theory,
we must define the {\it quasiparticle vacuum}
on which the vibrations are built.
The observed moments of inertia $\J$ of the yrast SD bands
smoothly increase in the A=190 region,
which suggests that the internal structure also smoothly changes as a
function of the frequency $\w$.
Therefore the {\it adiabatic representation}, in which the quasiparticle
operators are always defined with respect to the yrast state
$\ket{\w}$, is considered to be appropriate in this work.

In terms of quasiparticles, the Hamiltonian of
eq.(\ref{sp-potential})
can be diagonalized (by the general Bogoliubov transformation) as
\begin{equation}
h_{\rm s.p.}=\mbox{const.}
 + \sum_\mu \left( E_\mu a_\mu^\dagger a_\mu \right)
 + \sum_{\bar\mu} \left( E_{\bar\mu} a_{\bar\mu}^\dagger a_{\bar\mu}
 \right) \ ,
\end{equation}
with
\begin{equation}
a_\mu \ket{\w} = a_{\bar\mu} \ket{\w} = 0 \ ,
\end{equation}
where ($a_\mu^\dagger, a_{\bar\mu}^\dagger$) represent the
quasiparticles with signature $\alpha=(1/2,-1/2)$, respectively.
The excitation operators
of the RPA normal modes ${X_n^\alpha}^\dagger$ ($\alpha=0,1$)
are defined by
\begin{eqnarray}
{X_n^0}^\dagger &=& \sum_{\mu\bar\nu} \left\{
  \psi_n^0(\mu\bar\nu) a_\mu^\dagger a_{\bar\nu}^\dagger
 +\varphi_n^0(\mu\bar\nu) a_{\bar\nu} a_\mu \right\} \ ,\\
\label{normal-mode-0}
{X_n^1}^\dagger &=& \sum_{\mu < \nu} \left\{
  \psi_n^1(\mu\nu) a_\mu^\dagger a_\nu^\dagger
 +\varphi_n^1(\mu\nu) a_\nu a_\mu \right\}
 +\sum_{\bar{\mu} < \bar{\nu}} \left\{
  \psi_n^1(\bar{\mu}\bar{\nu}) a_{\bar\mu}^\dagger a_{\bar\nu}^\dagger
 +\varphi_n^1(\bar{\mu}\bar{\nu}) a_{\bar\nu} a_{\bar\mu} \right\} \ ,
\label{normal-mode-1}
\end{eqnarray}
where indices $n$ specify excited states and
$\psi_n^\alpha(\mu\nu)$ ($\varphi_n^\alpha(\mu\nu)$)
are the RPA forward (backward) amplitudes.
Quasiparticle-scattering terms such as $a_\mu^\dagger a_\nu$
are regarded as higher-order terms in the boson-expansion theory
and are neglected in the RPA\footnote{
In the following, the notation $[A,B]_{\rm RPA}$ means that we
neglect these higher order terms in calculating the
commutator between $A$ and $B$.}.

The equation of motion and the normalization condition
in the RPA theory,
\begin{eqnarray}
\left[ h_{\rm s.p.}+H_{\rm int}, {X_n^\alpha}^\dagger
\right]_{\rm RPA} &=& \hbar\Omega_n^\alpha {X_n^\alpha}^\dagger \ ,
\label{RPA-eom}\\
\left[ X_n^\alpha, {X_n^\alpha}^\dagger \right]_{\rm RPA} &=&
\delta_{nn'} \ ,
\label{RPA-norm}
\end{eqnarray}
are solved with the following multi-dimensional response functions:
\begin{equation}
S_{\rho\rho'}^\alpha (\Omega) =
\sum_{\gamma\delta} \left\{
 \frac{{R_\rho^\alpha(\gamma\delta)}^* R_{\rho'}^\alpha(\gamma\delta)}
      {E_\gamma+E_\delta-\hbar\Omega}
+\frac{R_\rho^\alpha(\gamma\delta) {R_{\rho'}^\alpha(\gamma\delta)}^*}
      {E_\gamma+E_\delta+\hbar\Omega}
 \right\} \ ,
\end{equation}
where $(\gamma\delta)=(\mu\bar\nu)$ for $\alpha=0$ states,
and $(\gamma\delta)=(\mu<\nu),(\bar\mu<\bar\nu)$
for $\alpha=1$ states.
The two-quasiparticle matrix elements
$R_\rho^\alpha(\gamma\delta)$ are defined by
$R_\rho^\alpha(\gamma\delta)\equiv\bra{\w}a_\delta a_\gamma
R_\rho^\alpha \ket{\w}$.
Let us denote the transition matrix elements between
the RPA excited states $\ket{n}$ and the yrast state as
\begin{equation}
t_\rho^\alpha (n) \equiv t_n\left[ R_\rho^\alpha \right]
\equiv \bra{\w} R_\rho^\alpha \ket{n}
= \bra{\w} \left[ R_\rho^\alpha, {X_n^\alpha}^\dagger \right] \ket{\w}
= \left[ R_\rho^\alpha, {X_n^\alpha}^\dagger \right]_{\rm RPA} \ .
\end{equation}
Then, the equation of motion (\ref{RPA-eom}) is equivalent to
\begin{equation}
t_\rho^\alpha (n) = \sum_{\rho'} \chi_\rho^\alpha
S_{\rho\rho'}^\alpha(\Omega) t_{\rho'}^\alpha (n) \ .
\label{RPA-eq}
\end{equation}
RPA solutions (eigen-energies) $\hbar\Omega_n$
are obtained by solving the equation,
\begin{equation}
\det \left( S_{\rho\rho'}^\alpha (\Omega) -
\frac{1}{\chi_\rho}\delta_{\rho\rho'} \right) = 0\ ,
\label{RPA_dispersion_eq}
\end{equation}
which corresponds to the condition that
eq.(\ref{RPA-eq}) has a non-trivial solution
($t_\rho^\alpha (n)\neq 0$).
Each RPA eigenstate is characterized by
the corresponding forward and backward amplitudes
which are calculated as
\begin{equation}
\psi_n^\alpha(\gamma\delta)
= \frac{\sum_\rho \chi_\rho^\alpha t_\rho^\alpha (n)
    R_\rho^\alpha(\gamma\delta)}
{E_\gamma  + E_{\delta} -\hbar\Omega_n}\ ,\quad
\varphi_n^\alpha(\gamma\delta)
= \frac{-\sum_\rho \chi_\rho^\alpha t_\rho^\alpha (n)
        {R_\rho^\alpha(\gamma\delta)}^*}
      {E_\gamma  + E_\delta +\hbar\Omega_n}\ ,
\end{equation}
and satisfies the normalization condition (\ref{RPA-norm}).
The transition matrix elements $\bra{\w} Q \ket{n}$
of any one-body operator $Q$ can be
expressed in terms of these amplitudes $\psi_n$ and $\varphi_n$.
\begin{eqnarray}
t_n\left[ Q \right] &\equiv& \bra{\w} Q \ket{n} \nonumber \\
&=& \sum_{\gamma\delta}\left\{
{Q(\gamma\delta)}^* \psi_n(\gamma\delta)
- Q(\gamma\delta) \varphi_n(\gamma\delta) \right\} \ .
\label{Q-amplitude}
\end{eqnarray}
The phase relation between the matrix elements $Q(\gamma\delta)$
and the amplitudes ($\psi_n(\gamma\delta),\varphi_n(\gamma\delta)$)
is very important, because it determines whether the transition matrix
element $t_n\left[ Q \right]$ is coherently enhanced or canceled out
after the summation in eq.(\ref{Q-amplitude}).
For instance, a collective quadrupole vibrational state has
a favorable phase relation for the quadrupole operators.
Therefore, it gives large matrix elements for the $E2$ operators,
while for the M1 operators,
the contributions are normally canceled out after the summation.

Finally we obtain a diagonal form of the total Hamiltonian
in the rotating frame by means of the RPA theory,
\begin{equation}
H' = h_{\rm s.p.} + H_{\rm int}
   \approx \mbox{const.}
 + \sum_{n,\alpha}
     \hbar\Omega_n^\alpha {X_n^\alpha}^\dagger X_n^\alpha \ .
\end{equation}
It is worth noting that since the effect of the cranking term on
the quasiparticles depends on rotational frequency,
the effects of Coriolis coupling
on the RPA eigenstates
are automatically taken into account.

%%%%%%%%%%%%% Details of Calculations %%%%%%%%%%%%%%%%%%%%%%%%%%%%%
\section{Details of calculations}
\label{sec: detail}

\subsection{The mean-field parameters and the improved quasiparticle
routhians}
\label{sec: detail-MF}

We adopt standard values for the parameters
$v_{ll}$ and $v_{ls}$\cite{BR85}
and use different values of the oscillator frequency $\omega_0$
for neutrons and protons
in the Nilsson potential (\ref{Nilsson})
in order to ensure equal root-mean-square
radii\cite{BK68}.
\begin{equation}
\label{oscillator_frequency}
\omega_0 \longrightarrow \left\{
\begin{array}{ll}
\left(\frac{2N}{A}\right)^{1/3} \omega_0\ ,
& \mbox{for neutrons} \ ,\\
\left(\frac{2Z}{A}\right)^{1/3} \omega_0\ ,
& \mbox{for protons} \ ,
\end{array} \right.
\end{equation}
where $\hbar\omega_0=41A^{-1/3}$MeV.

The quadrupole deformation $\epsilon$ is determined by
minimizing the total routhian surface (TRS),
and the strength for the monopole
pairing interaction $G$ is taken from the prescription
of Ref.\cite{Bra72} with the average pairing gap
$\tilde\Delta = 12A^{-1/2}$MeV and the cut-off parameter of
the pairing model space $\Lambda = 1.2\hbar\omega_0$.
In principle the pairing gaps ($\Delta_n,\Delta_p$)
and the chemical potentials ($\lambda_n,\lambda_p$)
should be calculated self-consistently
satisfying the usual BCS conditions at each rotational frequency:
\begin{eqnarray}
G_\tau \bra{\w}P_\tau \ket{\w} &=& \Delta_\tau\ ,
\label{gap_equation}\\
\bra{\w} N_\tau \ket{\w} &=& N (Z) \quad\mbox{ for } \tau=n(p)\ .
\label{number_equation}
\end{eqnarray}
However, the mean-field treatment of the pairing interaction
predicts a sudden collapse of the proton pairing gap at
$\hbar\w\approx0.3\mbox{MeV}$ and of the neutron gap
at $\hbar\w\approx0.5\mbox{MeV}$.
This transition causes a singular behavior in the moments of inertia
which is inconsistent with experimental observations.
It arises from the poor treatment of number conservation,
and such sudden transitions should not occur in a finite system
like the nucleus.
In this paper we have therefore adopted the following
phenomenological prescription for the pairing correlations at finite
frequency\cite{Wys90}:
\begin{equation}
\label{phenom_pairing}
\Delta_\tau(\omega) = \left\{
\begin{array}{ll}
\Delta_\tau(0) \left(1-\frac{1}{2}
\left( \frac{\omega}{\omega_c} \right)^2\right),
& \mbox{for }\omega < \omega_c, \\
\frac{1}{2}\Delta_\tau({0})
\left( \frac{\omega_c}{\omega} \right)^2,
& \mbox{for }\omega > \omega_c,
\end{array}
\right.\ .
\end{equation}
The chemical potentials are calculated with eq.(\ref{number_equation})
at each rotational frequency.
The parameters $\Delta(0)=0.8$ (0.6) MeV and
$\hbar\omega_c=0.5$ (0.3) MeV for neutrons (protons)
are used in common for $^{190,192,194}$Hg.

The quadrupole deformation $\epsilon=0.44$ is used in the calculations.
For simplicity, we assume the deformation to be constant with
rotational frequency, and neglect hexadecapole deformation.
The equilibrium deformation and pairing gaps have been determined at
$\w=0$, with the truncated pairing model space $\Lambda=1.2\hbar\omega_0$.
Then, the pairing force strengths $G_\tau$ are adjusted so as to
reproduce the pairing gap of eq.(\ref{phenom_pairing})
in the whole model space.

The experiments\cite{Cro94,Cro95} have reported a sharp rise of $\J$
moments of inertia for the yrast SD band in \hgo\
at $\hbar\w\approx 0.4 \mbox{MeV}$.
This rise was reproduced in the cranked Woods-Saxon
calculations\cite{Dri91} and results from a crossing
between the yrast band and the aligned
$\nu(j_{15/2})^2$ band, however,
the predicted crossing frequency was
lower ($\hbar\w\approx 0.3\mbox{MeV}$)
than in the experiment.
Our Nilsson potential without the additional term (\ref{additional-term})
indicates the same disagreement.
In order to demonstrate the effects of the term $h_{\rm add}$
on the routhians, we present in Fig.~\ref{hg190_routh_comp}
the quasiparticle routhians for \hgo\ with $h_{\rm add}$,
without $h_{\rm add}$,
and for the standard Woods-Saxon potential
($\beta_2=0.465$, $\beta_4=0.055$).
By including,
$h_{\rm add}$,
the correct frequency is reproduced.
This term affects the proton routhians: for example,
the alignment of the intruder $\pi j_{15/2} (\alpha=-1/2)$ orbit is
predicted to be $i\approx 6.5\hbar$ without $h_{\rm add}$
and this orbit becomes the lowest at $\hbar\w \geq 0.37\mbox{MeV}$.
The alignment is significantly reduced ($i\approx 4\hbar$)
with $h_{\rm add}$.
The behavior of high-$N$ intruder orbits in the proton routhians
are similar to those in the Woods-Saxon potential.
It is worth noting that
the conventional renormalization in the Nilsson potential scales the
rotational frequency for all orbits,
while eq.(\ref{additional-term}) renormalizes alignment in a
different way depending on the spurious effect on each orbit.

\subsection{The residual interactions and the RPA}
\label{sec: detail-RPA}

We adopt the following operators as $R_\rho^\alpha$
in the residual interactions (\ref{residual-int}).
\begin{equation}
\label{operator_R}
\begin{array}{lllllll}
P_+ & P_- & Q_{20}^0 & Q_{21}^\alpha & Q_{22}^\alpha & &
 \mbox{for positive-parity states}\ ,\\
Q_{30}^1 & Q_{31}^\alpha & Q_{32}^\alpha & Q_{33}^\alpha
& \tilde\tau_3 Q_{10}^1 & \tilde\tau_3 Q_{11}^\alpha
& \mbox{for negative-parity states}\ ,
\end{array}
\end{equation}
where $\tilde\tau_3 = \tau_3 -(N-Z)/A$ which is needed to guarantee the
translational invariance.
Here, the operators $Q_{\lambda K}^\alpha$ are defined by
eq.(\ref{multipole-op}) in the doubly-stretched coordinates,
and $P_\pm$ are defined by
\begin{eqnarray}
P_+ &=& \frac{1}{\sqrt{2}}\left(\tilde{P} + \tilde{P^\dagger}\right)\ ,\\
P_- &=& \frac{i}{\sqrt{2}}\left(\tilde{P} - \tilde{P^\dagger}\right)\ ,
\end{eqnarray}
where $\tilde{P}=P-\bra{\w}P\ket{\w}$.
Note that the $K=0$ quadrupole (octupole) operator $Q_{20}$ ($Q_{30}$)
has a unique signature $\alpha=0$ ($\alpha=1$),
which corresponds to the fact that $K=0$ bands
have no signature partners.

Since we use the different oscillator frequency $\omega_0$
for neutrons and protons
in the Nilsson potential (see eq.(\ref{oscillator_frequency})),
we use the following modified
doubly-stretched multipole operators for the isoscalar channels:
\begin{equation}
Q_{\lambda K}^\alpha \longrightarrow \left\{
\begin{array}{ll}
\left(\frac{2N}{A}\right)^{2/3} Q_{\lambda K}^\alpha\ ,
& \mbox{for neutrons}\ ,\\
\left(\frac{2Z}{A}\right)^{2/3} Q_{\lambda K}^\alpha\ ,
& \mbox{for protons}\ .
\end{array}
\right.
\end{equation}
This was originally proposed by Baranger and Kumar\cite{BK68} for
quadrupole operators.
Recently Sakamoto\cite{Sak93} has generalized it
for an arbitrary multipole operator and
proved that by means of this scaling
the translational symmetry is restored in the limit of the
harmonic-oscillator potential.
In addition, for the collective RPA solutions this treatment makes
the transition amplitudes of the electric operators approximately
$Z/A$ of those of the mass operators, in the same way as in
the case of the static quadrupole moments\cite{SM95}.

We use the pairing force strengths $G_\tau$
reproducing the pairing gaps of eq.(\ref{phenom_pairing}).
For the isovector dipole coupling strengths,
we adopt the standard values in Ref.\cite{BM75},
\begin{equation}
\chi_{1K}=-\frac{\pi V_1}{A\langle (r^2)^{''} \rangle_0}\ ,
\end{equation}
with
$A\langle (r^2)^{''} \rangle_0 = \langle \sum_k^A (r_k^2)^{''} \rangle_0$
and $V_1 = 130$MeV.
The self-consistent values for the coupling strengths
$\chi_{\lambda K}$ of
the isoscalar quadrupole and octupole interactions
can be obtained for the
case of the anisotropic harmonic-oscillator
potential\cite{SK89,Sak93}:
\begin{eqnarray}
\chi_{2K}^{\rm HO} &=&
\frac{4\pi M\omega_0^2}{5A\langle(r^2)^{''}\rangle}\ ,\\
\chi_{3K}^{\rm HO} &=&
{4\pi \over 7} M\omega^2_0
\biggl\{A\langle (r^4)^{''}
 \rangle +
    {2 \over 7}(4 - K^2) A\langle (r^4P_2)^{''}\rangle \nonumber\\
&+& {1 \over 84}(K^2(7K^2 - 67) + 72) A\langle (r^4P_4)^{''}
 \rangle \biggr\}^{-1}\ ,
\end{eqnarray}
with
\begin{equation}
A\langle (r^n P_\l )\rangle \equiv \left(\frac{2N}{A}\right)^{2/3}
\langle \sum_k^N (r_k)^n P_\l \rangle_0
+ \left(\frac{2Z}{A}\right)^{2/3}
\langle \sum_k^Z (r_k)^n P_\l \rangle_0\ .
\end{equation}

A large model space has been used for solving the coupled RPA equations,
including seven major shells with
$N_{\rm osc}=3\sim 9$ $(2\sim 8)$ for neutrons (protons) in the
calculations of positive-parity states,
and nine major shells with
$N_{\rm osc}=2\sim 10$ $(1\sim 9)$ for the negative-parity states.
The mesh of the rotational frequency for the calculations has been
chosen as ${\it\Delta}\hbar\w=0.01 \mbox{MeV}$ which is enough to
discuss the properties of band crossing and Coriolis couplings.

Since our mean-field potential is not the simple harmonic oscillator,
we use scaling factors $f_\lambda$ as
\begin{equation}
\chi_{\lambda K} = f_\lambda \cdot \chi_{\lambda K}^{\rm HO}\ ,
\label{coupling-strength}
\end{equation}
for the isoscalar interactions with $\lambda=2$ and 3.
These factors are determined by the theoretical
and experimental requirements:
As for the octupole interactions,
we have the experimental routhians
for the lowest octupole vibrational state in SD \hgo\cite{Cro95}.
We assume the common factor $f_3$ for all $K$-values and
fix it so as to reproduce these experimental data.
In this case $f_3=1$ can nicely reproduce the experimental
routhians\footnote{
This value depends on the treatment of the pairing gaps at finite
frequency.
If we use constant pairing gaps against $\w$ we get the best value
$f_3=1.05$.},
and we use the same value for \hgt\ and \hgf.
For the quadrupole interactions,
we determine it so as to reproduce the zero-frequency (Nambu-Goldstone)
mode for $K=1$ at $\w=0$ and use the same value for $K=0$ and 2.
$f_2=1.007$, 1.005, and 1.005 are obtained for
\hgo, \hgt, and \hgf, respectively, by using the adopted model space.
The fact that these values of $f_\lambda$ are close to unity indicates
that the size of the adopted model space is large enough.

According to systematic RPA calculations for the
low-frequency $\beta$, $\gamma$, and octupole states in medium-heavy
deformed nuclei,
we have found that the values of $f_\lambda$
reproducing the experimental data are very close to unity
for the Nambu-Goldstone mode,
the $\gamma$ and octupole vibrational states.
On the other hand, those values are quite different from unity
for the $\beta$ vibrational states.
This may be associated with the simplicity of the monopole pairing
interaction.
Since we can not find the realistic force strength $\chi_{20}$ for
SD states,
we do not discuss the property of the $\beta$ vibrations in this paper.

%%%%%%%%%%%%%%%% The Calculated Results %%%%%%%%%%%%%%%%%%%%%%%%%%%%
\section{The results of numerical calculations}
\label{sec: results}

\subsection{Quasiparticle routhians}

In this section we present calculated quasiparticle routhians in the
improved cranked Nilsson potential
and discuss their characteristic feature.
In Fig.~\ref{neutron_routhians} we compare
the neutron quasiparticle routhians
for $^{190,192,194}$Hg.
The proton routhians of \hgo\ are shown above in
Fig.~\ref{hg190_routh_comp}
and are almost identical for \hgt\ and \hgf.

The calculations show the strong interaction strength between
the $\pi($[642 5/2]$)^2$ configuration
(for simplicity we denote these orbits by $\pi 6_1$ and
$\pi 6_2$ in the following)
and the yrast configuration which may contribute to the
smooth increase of the yrast $\J$ moments of inertia.
On the other hand, the interaction of $\nu$[761 3/2]
orbits ($\nu 7_1$ and $\nu 7_2$ in the following)
strongly depends on the chemical potential (neutron number):
The interaction is strongest in \hgf, and weakest in \hgo.
This is qualitatively consistent with the experimental observation
of the yrast $\J$ moments of inertia
and the experimental quasiparticle routhians in
$^{191,193}$Hg\cite{Joy94,Car95}.

The characteristic features of the high-$N$ intruder orbits are
similar to those of the Woods-Saxon potential,
except the alignments of $\nu 7_1$ and $\nu 7_2$ orbits
which are, respectively, $i\approx 3\hbar$ and $2\hbar$
in ours while $i\approx 4\hbar$ and $3\hbar$ in Woods-Saxon's.
This results in the different crossing frequency between
the ground band and the $\nu (j_{15/2})^2$ band, as discussed
in section \ref{sec: detail-MF}.
The observed crossing in \hgo\ and the quasiparticle routhians
in $^{191,193}$Hg seem to favor our results.
There are some other minor differences
concerning the position of each orbit
in the Nilsson and in the Woods-Saxon potential.
However, these differences do not seriously affect our main conclusions
because the collective RPA solutions are not sensitive to
the details of each orbit.

%%%%%%%%%%%%%%%% The results for octupole vibrations %%%%%%%%%%%%%%%%%%%%
\subsection{The octupole vibrations}
\label{sec: octupole-results}

Here, we discuss the negative-parity excitations
in SD $^{190,192,194}$Hg.
We have solved the RPA dispersion equation (\ref{RPA_dispersion_eq})
and have obtained all low-lying solutions ($E'_x \leq  2$MeV).
The excitation energies and the $B(E3)$ values
calculated at $\w=0$ are listed in Table~\ref{oct-E-E3}.
This result shows that $K=2$ octupole states are the lowest
for these Hg isotopes,
which is consistent with our previous results\cite{NMM93,Miz93}.
The $B(E3; 0^+ \rightarrow 3^-,K)$ are calculated
by using the strong coupling scheme\cite{BM75} neglecting effects
of the Coriolis force.
Absolute values of $B(E3)$s cannot be taken seriously because
they depend on the adopted model space and
are very sensitive to the octupole coupling strengths $\chi_{3K}$:
For instance, if we use $f_3=1.05$ instead of $f_3=1$
in eq.(\ref{coupling-strength}),
the $B(E3)$ increase by about factor of two
while the reduction of their excitation energy is about 15\%.
In addition, the effects of the Coriolis coupling
tend to concentrate the $B(E3)$ strengths onto the lowest octupole
states\cite{NV70}.

At $\w=0$, the lowest $K=2$ octupole states exhibit almost identical
properties in $^{190,192,194}$Hg.
However they show different behavior as functions of $\w$ as
shown in Figs.~\ref{oct_hg190}, \ref{oct_hg192}, and
\ref{oct_hg194}, respectively.
All RPA solutions,
including non-collective solutions as well as collective vibrational
ones,
are presented in these figures.
The size of the circle on the plot indicates the magnitude of the $E3$
transition amplitudes
between an RPA solution and the yrast state.

The $(K,\alpha)=(2,1)$ octupole state
in \hgo\ has significant Coriolis mixing
and the octupole phonon is aligned along the rotational axis at higher
frequency.
This is caused by the relatively close energy spacing
between the $K=2$ and
the $K=0, 1$ octupole states in this nucleus.
These low-$K$ members of the octupole multiplet are calculated to lie much
higher in \hgt\ and \hgf, which reduces the Coriolis mixing
in these nuclei.
As a result of these phonon alignments,
the experimental routhians for Band 2 in \hgo\ are
nicely reproduced by the lowest $\alpha=1$ octupole state.
It should be emphasized that although the excitation energy
at one frequency point can be obtained by adjusting
the octupole-force strengths,
the agreement over the whole frequency region is not trivial.

Since there is no $K=0$ octupole state in the signature $\alpha=0$ sector,
the Coriolis mixing is much weaker for the lowest
$(K,\alpha)=(2,0)$ octupole state.
The calculation predicts that this state is crossed by the negative-parity
two-quasiparticle band $\nu(7_1\otimes[642\ 3/2])_{\alpha=0}$ at
$\hbar\w\approx 0.27\mbox{MeV}$.

In \hgt, the same kind of crossing is seen for both signature
partners of the $K=2$ octupole bands.
We can clearly see, for the lowest excited state in each signature sector,
the transition of the internal structure
from collective octupole states (large circles in Fig.~\ref{oct_hg192})
to non-collective two-quasineutrons (small circles).
The two-quasineutron configurations which cross the octupole vibrational
bands are $7_1\otimes[642\ 3/2](\alpha=-1/2)$ for $\alpha=1$
and $7_1\otimes[642\ 3/2](\alpha=1/2)$ for $\alpha=0$.
The crossing frequency is lower for the $\alpha=1$ band
due to signature splitting of the $\nu$[642 3/2] orbits.

In contrast to $^{190,192}$Hg,
the $K=2$ octupole bands in \hgf\ indicate
neither the signature splitting nor the crossings.
The routhians are very smooth up to the highest frequency.
This is because the neutron orbits $7_1$ and $7_2$ have a
``hole'' character
and their interaction strengths with the negative-energy orbits
become larger with increasing neutron numbers
(see Fig.~\ref{neutron_routhians}).
Therefore these orbits go to higher energy
and the energies of the two-quasiparticle bands
$\nu(7_1\otimes[642\ 3/2])$
never come lower than the $K=2$ octupole bands
even at the highest frequency.

These properties of the $K=2$ octupole vibrations come from the
effects of the Coriolis force
and from the chemical-potential dependence of the aligned
two-quasiparticle bands.
In order to reproduce these rich properties
of the collective vibrations at finite frequency,
a microscopic model, which can describe
the interplay between the Coriolis force and the
correlations of shape fluctuations, is needed.

%%%%%%%%%%%%%%%% The results for gamma vibrations %%%%%%%%%%%%%%%%%%%%
\subsection{The $\gamma$ vibrations}
\label{sec: gamma-results}

In this section we present results for the
$\gamma$-vibrational states built on the
SD yrast band.
As mentioned in section \ref{sec: detail-RPA},
we do not discuss the property of the $\beta$ band
since it is difficult to determine a reliable value of the coupling strength
$\chi_{20}$ for the K=0 channel of the quadrupole interaction.

The properties of $\gamma$ bands at $\w=0$ are listed in
Table~\ref{gamma-E-E2}.
The excitation energies of $\gamma$ vibrations are predicted
to be higher than the $K=2$ octupole vibrations by 200--350 keV.
It is known that calculations using the full model space
considerably overestimate the $B(E2)$ values.
In Ref.\cite{SM95}, it has been shown that the three $N_{\rm osc}$-shells
calculation reproduces the experimental values very well.
If we use the model space $N_{\rm osc}=5\sim 7$ ($4\sim 6$) for neutrons
(protons), then the $B(E2)$ values in the table
decrease by about factor 1/3.
The collectivity of the $\gamma$ vibrations turns out to be very weak in
these SD nuclei.

Figs.~\ref{quad_hg190}, \ref{quad_hg192}, and
\ref{quad_hg194} illustrate the excitation energy of
$\gamma$ vibrations as functions of the rotational frequency
for \hgo, \hgt, and \hgf, respectively.
The unperturbed two-quasiparticle routhians are also depicted by solid
(neutrons) and dashed (protons) lines.
Since the $K$ quantum number is not a conserved quantity at finite
rotational frequency,
we have defined the solutions with the large $K=2$ $E2$ transition
amplitude as the $\gamma$ vibrations.
As seen in the figure,
they lose their vibrational character
by successive crossings with many
two-quasiparticle bands and become the dominant
two-quasiparticle states at high frequency.
The reduction of collectivity is more rapid for the $\alpha=0$ $\gamma$
vibrations, because the two-quasiparticle states come down more quickly
in the $\alpha=0$ sector.
Similar crossings occur for the $K=2$ octupole bands
in \hgt\ (see Fig.~\ref{oct_hg192}),
however, the crossing frequency is much higher than that of the
$\gamma$ bands.
This is because the excitation energies of the octupole bands
are relatively lower than those of the $\gamma$ bands.
The predicted properties of $\gamma$ vibrations are different from
those in Ref.\cite{Gir92}.

In the frequency region ($0.15\leq\hbar\w\leq 0.4\mbox{MeV}$) where the
excited SD bands are observed in experiments,
the $\gamma$ bands are predicted to be higher than both
the $K=2$ octupole bands and the lowest two-quasiparticle states.
Therefore the experimental observation of the $\gamma$ vibrations
is expected to be more difficult than that of the octupole bands.

%%%%%%%%%%%%%% Comparison with experiments %%%%%%%%%%%%%%%%%%%%%%%
\section{Comparison with experimental data}
\label{sec: comparison}

In this section, we compare the results obtained in the previous section
with the available experimental data for the excited SD bands in
$^{190,192,194}$Hg.
The routhians relative to the yrast SD band
have been observed only for Band 2 in \hgo\ and
the comparison with our calculated routhians has been done in the
section \ref{sec: octupole-results}.
The excitation energies of the other bands are not known.
Therefore, in order to compare our theory with experimental data,
we have calculated the dynamic moments of inertia $\J$.

The $\J$ of the excited bands are calculated as
\begin{equation}
\J(\omega) = \J_0(\omega) + \frac{di}{d\omega}
           = \J_0(\omega) - \frac{d^2 E'_x}{d\omega^2} \ ,
\label{J2-inertia}
\end{equation}
where $\J_0$ denotes the dynamic moments of inertia for the yrast SD
bands (RPA vacuum), and $i$ and $E'_x$ are the calculated alignments and
routhians relative to the yrast band, respectively.
The $\J_0$ values of the yrast SD bands are taken from the experiments
and approximated by the Harris expansion,
\begin{equation}
\J_0(\omega) = J_0 + 3 J_1 \omega^2 + 5 J_2 \omega^4 \ .
\label{Harris}
\end{equation}
It is known that the effect of pairing fluctuations is important in
reproducing the moments of inertia at high spin.
However, since our model provides us with relative quantities
(excitation energy, alignment, etc.)
between the excited bands and the yrast band,
it is not critical if we neglect the pairing fluctuations.
In other words, the fluctuations are included in the experimental $\J_0$
of eq.(\ref{J2-inertia}).

The lower the excitation energy of an excited band
relative to the yrast SD band,
the more strongly will it be populated.
In experiments, the SD bands are populated at high frequency,
thus, it is the excitation energy in the feeding region
at high frequency that is relevant in this
problem.
We list in Table~\ref{Ex} the calculated excitation energies of
the low-lying positive- and negative-parity states
at $\hbar\w=0.4\mbox{MeV}$.

In \hgo\ three excited SD bands (Bands 2, 3 and 4) have been
observed\cite{Cro94,Cro95,Wil95}.
Band 2 has been assigned as the lowest octupole band\cite{Cro94,Cro95}
because of its strong decays into the yrast SD band.
According to our calculations,
in addition to this octupole band ($\alpha=1$),
the aligned two-quasineutron bands come down at high frequency.
We assign Band 4 at high frequency
as the $\nu (7_1 \otimes [642\ 3/2])_{\alpha=0}$
because this negative-parity two-quasineutron state is crossed by the
$\alpha=0$ octupole band at $\hbar\w\approx 0.26\mbox{MeV}$ which
may correspond to the observed sharp rise of $\J$ at $\hbar\w\approx
0.23\mbox{MeV}$ (Fig.~\ref{oct_hg190}).
The positive-parity $\nu (7_1 \otimes 7_2)_{\alpha=0}$ state is also
relatively low-lying at high frequency.
Since this band does not show any crossing
at $\hbar\w > 0.12\mbox{MeV}$ in the calculations,
this may be a good candidate for Band 3 (Fig.~\ref{quad_hg190}).

In \hgt\, two excited SD bands (Bands 2 and 3) have been
observed\cite{Fal95} and both bands exhibit a bump in $\J$
at $\hbar\w\approx 0.3$ (Band 2) and $0.33\mbox{MeV}$ (Band 3).
We assume these bands correspond to
$\nu(7_1\otimes [642\ 3/2])_{\alpha=0,1}$ at high frequency.
This two-quasineutron configuration for Band 2 is the same as that
suggested in Ref.\cite{Fal95}.
However our theory predicts a different scenario at low spin:
This band is crossed by the octupole band ($\alpha=1$)
at $\hbar\w\approx 0.3\mbox{MeV}$.
Thus, Band 2 is interpreted as an $\alpha=1$ octupole vibrational
band in the low-frequency region ($\hbar\w<0.3\mbox{MeV}$).
In the same way, the bump in $\J$ in Band 3
is interpreted as a crossing between
$\nu(7_1\otimes [642\ 3/2])_{\alpha=0}$ and the $\alpha=0$ octupole
vibrational band (Fig.~\ref{oct_hg192}).

For high frequencies, the positive-parity
$\nu(7_1 \otimes [512\ 5/2])$ state is
calculated to lie almost at the same energy as the lowest $\alpha=0$
negative-parity state.
However no crossing is predicted for the $\alpha=1$ state at
$\hbar\w > 0.15\mbox{MeV}$
but many crossings are predicted for the $\alpha=0$ state
(Fig.~\ref{quad_hg192}).
Both properties are incompatible with the observed features.

In {\hgf}, two excited SD bands (Bands 2 and 3) have been
observed\cite{Ril90,Ced95}.
In contrast to \hgt, the observed dynamic moments of inertia $\J$
do not show any singular behavior and are more or less similar
to those of the yrast band.
Bands 2 and 3 have been interpreted as signature partners
because the $\gamma$-ray energies of Band 3 are observed to lie
mid-way between those of Band 2
and furthermore the bands have similar intensity\cite{Ril90}.
{}From these observations and the excitation energies listed in
Table~\ref{Ex},
we assume that both bands correspond to $K=2$ octupole vibrations
($\alpha=0$, 1), which are calculated to be
the lowest excited states (Fig.~\ref{oct_hg194}).
Any other assignment faces serious difficulties:
(i) The positive-parity two-quasiparticle configurations listed in
Table~\ref{Ex} have no signature partners.
(ii) The other low-lying two-quasiparticle states occupy
$\nu 7_1$ or $\pi 6_1$ orbits.
Now the increase in $\J$ for the yrast SD band is partially attributed
to the alignment of these high-$j$ intruder orbits and,
since the blocking effect of the quasiparticles prevents
any alignment due to band crossings involving these orbits,
the lack of alignment should produce an $\J$ curve quite different from
those of the yrast SD band.
(iii) The configuration $\nu ([512\ 5/2]\otimes [624\ 9/2])$
suggested in Ref.\cite{Ril90} has the problem with its magnetic
property,
which has been recently pointed out in Ref.\cite{SRA95}.
If this configuration is the $K^\pi=7^-$, then strong M1 transitions
between the signature partners should have been observed.
The energy of the $K^\pi=2^-$ configuration is certainly lowered by
octupole correlations.
In our calculations, however, this configuration accounts for only
20\% of all components constituting the octupole vibration.
The $\gamma$ vibrations are calculated to be much higher and crossed by
several two-quasiparticle bands (Fig.~\ref{quad_hg194}).
Therefore,
we believe the octupole vibration is the best candidate\footnote{
The signature for Bands 2 and 3 is determined by
following the spin assignment in Ref.\cite{Ril90}.}.

Assuming the above configurations,
the dynamic moments of inertia $\J$ are calculated with
eq.(\ref{J2-inertia}), and compared with the experimental data
(Fig.~\ref{J2}).
In \hgo,
the characteristic features are well reproduced for Bands 2 and 4;
the constant $\J$ of Band 2 (the $\alpha=1$ octupole vibration), and
the bump of Band 4 (the crossing between the $\alpha=0$ octupole vibration
and the aligned two-quasineutron band) are reproduced
although the crossing frequency is
smaller in the experiment.
For Band 3, the high $\J$ values
at low spin are well accounted for by the alignment-gain of
the two-quasineutron state.
However the calculation predicts the lack of alignment due to the blocking
of $N=7$ orbits
at $\hbar\w>0.25\mbox{MeV}$, which makes the $\J$ smaller than those of
the yrast band.

In \hgt,
the bumps of $\J$ are nicely reproduced in the calculations, which
correspond to the crossings between $K=2$ octupole vibrations and the
aligned two-quasineutron bands in each signature partner.
The alignment gain ${\it \Delta}i$ before and after crossing for Band 2
is ${\it\Delta} i\approx 2\hbar$ which is comparable to the experimental
value ${\it\Delta} i_{\rm exp}\approx 2.6\hbar$\cite{Fal95}.

The agreement is less satisfactory in \hgf.
The calculated $\J$ are lower than the experimental data for
$0.2\leq\hbar\w\leq 0.35\mbox{MeV}$ (similar disagreement can be
seen for Band 3 in \hgt).
This effect comes from the blocking effect mentioned above,
associated with the $\nu 7_1$, $\nu 7_2$, $\pi 6_1$ and $\pi 6_2$ orbits.
In the RPA (Tamm-Dancoff) theory (neglecting the backward amplitudes),
the octupole vibrations are described by superposition of
two-quasiparticle excitations,
\begin{equation}
\ket{\rm oct.vib.} =
\sum_{\gamma\delta} \psi(\gamma\delta)\ket{\gamma\delta}\ ,
\end{equation}
where $\ket{\gamma\delta}=a_\gamma^\dagger a_\delta^\dagger \ket{\w}$.
Some of these components $\ket{\gamma\delta}$ associated with
the particular orbits ($\nu 7_1$, $\nu7_2$, $\pi 6_1$ and $\pi6_2$)
show significant lack of alignment.
However, if the octupole vibrations are collective enough,
the amplitudes $\psi(\gamma\delta)$ are distributed over many
two-quasiparticle excitations $\ket{\gamma\delta}$.
Thus, each amplitude becomes small
and blocking effects may be canceled.

In order to demonstrate this ``smearing'' effect of collective states,
we use a slightly stronger octupole
force, $f_3=1.05$ in eq.(\ref{coupling-strength}), and carry out the same
calculations for \hgf.
The results are shown in Fig.~\ref{J2_hg194}.
The higher coupling strengths make the octupole vibrations more
collective and the experimental data are better reproduced.
Perhaps the collectivity of these octupole vibrations
was underestimated in the calculations with $f_3=1$.

Finally we should mention the decays from the octupole bands to the yrast SD
band.
We have assigned all observed excited SD bands (except Band 3 in \hgo) as
octupole vibrational bands (at least in the low-spin region).
However, strong dipole decays into the yrast band
have been observed only for Band 2 in \hgo.
Although this seems to contradict our proposals, in fact
our calculations provide us with a qualitative answer.

Let us discuss the relative
$B(E1; {\rm oct}\rightarrow {\rm yrast})$ values.
Using the $E1$ recoil charge ($-Ze/A$ for neutrons and $Ne/A$ for protons),
then the $B(E1)$ values at $\hbar\w=0.25\mbox{MeV}$ are calculated to be small
for all the $K=2$ octupole bands except for the $\alpha=1$ (Band 2) in \hgo:
With the scaling factors $f_3=1\sim 1.08$ in eq.(\ref{coupling-strength}),
the calculation suggests
$B(E1)\approx 10^{-7}$ W.u. for the $(K,\alpha)=(2,0)$ octupole bands,
and $B(E1)\approx 10^{-8}\sim 10^{-6}$ W.u.
for the $(K,\alpha)=(2,1)$ bands.
The $B(E1)$ for Band 2 in \hgo\ is predicted to be larger than these values
by 1 -- 2 orders of magnitude, $B(E1)\approx 10^{-6}\sim 10^{-4}$W.u.
Although the absolute values are very sensitive to the
parameters used in the calculation,
the $E1$ strengths of Band 2 in \hgo\ are always much
larger than those for the other bands.

To clarify the reason for this $E1$ enhancement in this particular band,
we display the $E3$ amplitudes ($K=0$, 1, 2 and 3) of these octupole states
as functions of frequency in Fig.~\ref{E3_amplitudes}.
As mentioned in section ~\ref{sec: octupole-results},
the Coriolis mixing is completely different between
Band 2 in \hgo\ and the others:
The former has significant Coriolis mixing at finite frequency
while the latter retains the dominant $K=2$ character up to very high spin.
Since the $K=2$ octupole components can not carry any $E1$ strength,
the strong $E1$ transition amplitudes come from Coriolis coupling,
namely the mixing of the $K=0$ and 1 octupole components.
Therefore, the observed decay property does not contradict our
interpretation.

%%%%%%%%%%%%%% Conclusion %%%%%%%%%%%%%%%%%%%%%%%%%%%%%%%%%%%%%%%
\section{Conclusions}
\label{sec: conclusion}

The microscopic structure of the $\gamma$ and the octupole vibrations
built on the SD yrast bands
in $^{190,192,194}$Hg
were investigated with
the RPA based on the cranked shell model.
The $K=2$ octupole vibrations are predicted to lie lowest.
To reproduce the characteristic features of
the experimental data it was essential to
include octupole correlations and the effect of
rapid rotation explicitly.
{}From the calculations, we assigned the following configurations to
the observed excited bands:\newline

\begin{tabular}{lll}
\hgo & Band 2 :&
the rotationally-aligned $\alpha=1$ octupole vibration.\\
     & Band 3 :&
the two-quasineutron band $\nu(7_1 \otimes 7_2)$.\\
     & Band 4 :&
the $(K,\alpha)=(2,0)$ octupole vibration at low spin,\\
     &          &
the two-quasineutron band $\nu(7_1 \otimes [642\ 3/2])_{\alpha=0}$
at high spin.\\
\hgt & Band 2 :&
the $(K,\alpha)=(2,1)$ octupole vibration at low spin,\\
     &          &
the two-quasineutron band $\nu(7_1 \otimes [642\ 3/2])_{\alpha=1}$
at high spin.\\
     & Band 3 :&
the $(K,\alpha)=(2,0)$ octupole vibration at low spin,\\
     &          &
the two-quasineutron band
$\nu(7_1 \otimes [642\ 3/2])_{\alpha=0}$ at high spin.\\
\hgf & Band 2 :&
the $(K,\alpha)=(2,0)$ octupole vibration.\\
     & Band 3 :&
the $(K,\alpha)=(2,1)$ octupole vibration.
\end{tabular}

\noindent
With these assignments, most of the experimentally observed features were
well accounted for in our theoretical calculations.

The Coriolis force makes the lowest octupole state
in \hgo\ align along the rotational axis,
while this effect is predicted to be very weak
for other octupole states.
This is due to the relatively low excitation energy of the $K=0$ ($\alpha=1$)
octupole state in \hgo,
in which the close spacing in energy of the octupole multiplet makes
the Coriolis mixing easier.
This aligned octupole phonon in \hgo\ reproduces the observed
behavior for Band 2.

Our interpretation for the excited SD bands in \hgt\ solves
a puzzle mentioned in Ref.~\cite{Fal95} in which
Band 2 was assigned as the two-quasineutron excitation
$\nu (7_3\otimes [642\ 3/2])$.
The bump in the  $\J$ curve was considered to be
associated with a crossing between
the $\nu 7_1$ and $\nu$[512 5/2] orbits.
According to this assignment,
we expect similar properties for the observed crossing
in \hgt\ and $^{193}$Hg, and
the difference of crossing frequencies and alignment gains was a puzzle.
This is no longer a puzzle in our interpretation
because the microscopic structure of
Band 2 is the octupole vibration (before the crossing).
Due to the correlation-energy gains,
the excitation energies of the octupole vibrations should
be lower than the unperturbed two-quasiparticle states.
Therefore it is natural that the observed crossing frequency is
larger than the one predicted by the quasiparticle routhians
without the octupole correlations.

Our interpretation also solves some difficulties in \hgf:
The smooth $\J$ behavior of Bands 2 and 3 can be explained by the
``smearing'' effect of the collective states.
The non-observation of
the expected strong M1 transitions
between Bands 2 and 3\cite{SRA95}
is solved by substituting the $K=2$ octupole vibrations
for the two-quasineutron states $\nu ([512\ 5/2]\otimes [624\ 9/2])$,
because the octupole correlations lower the $K=2$ configurations
and the summation of many two-quasiparticle ($M1$) matrix elements
may be destructive (see discussion below eq.(\ref{Q-amplitude})).

The enhanced $E1$ transitions from the octupole states to the yrast SD band
are expected only for Band 2 in \hgo.  This comes about
because the other octupole states do not have strong Coriolis mixing
and keep their $K=2$ character even at high frequency.
This agrees with experimental observations.

Although most of the observed properties are explained by
our calculations,
there remain some unsolved problems in \hgo\ and \hgt.
For \hgo,
according to the calculations with constant pairing gaps reported in
Ref.\cite{Nak95_2},
it is suggested that Band 4 may correspond to the $(K,\alpha)=(1,0)$
octupole band which is predicted to be crossed by the two-quasineutron
band $\nu(7_1\otimes [642\ 3/2])_{\alpha=0}$
at $\hbar\w\approx 0.21\mbox{MeV}$.
Because of the phenomenological treatment for the pairing gaps
at finite frequency,
it is difficult to deny this possibility.
The experimental intensity of Band 3 raises another ambiguity:
Since it is much weaker than Bands 2 and 4,
it might be associated with a higher-lying configurations\cite{Wil95}.
For \hgt,
our calculations predict no signature splitting for the lowest octupole
bands at $\hbar\w\leq 0.25\mbox{MeV}$.
Therefore one may expect $\gamma$-ray energies
typical of the signature-partner pair
for Bands 2 and 3 similar to that in \hgf,
which is different from what is observed\cite{Fal95}.
Improvement of the pairing interactions (fluctuations, quadrupole
pairing) might solve these problems as well as
enable us to perform reliable RPA calculations
for $\beta$ vibrations.

Theoretical study of octupole vibrations carrying large $E1$
strengths would be of great interest,
because this could offer direct experimental evidence.
An improved version of calculations for $E1$ strengths
of high-spin octupole bands are in progress,
taking into account the restoration of translational and Galilean
invariance.
The $K=0$ octupole vibration in $^{152}$Dy has been predicted in
Ref.\cite{Nak95} and its decay into the yrast
band has been suggested \cite{Dag95}.
Strong $E1$ transition probabilities have been suggested by
Skalski\cite{Ska94} for $K=0$ octupole states
in the A=190 region.
Therefore, the search for low-lying low-$K$ octupole vibrations
is an important subject for the future.

%%%%%%%%%%%%%%%%%%%%%%%%% Acknowledgments %%%%%%%%%%%%%%%%%%%%%%%%
\acknowledgements

We would like to acknowledge W.~Nazarewicz for discussions and suggestions
for this paper.
One of authors (T.N.) also thank B.~Crowell, P.~Fallon,
J.F.~Sharpey-Schafer, J.~Skalski and A.N.~Wilson for valuable discussions.
Three of us (T.N., K.M. and Y.R.S.) thank the Institute for Nuclear Theory
at the University of Washington for its hospitality and the Department of
Energy for partial support during the completion of this work.

%%%%%%%%%%%%%%%%%%%%%%%%% References %%%%%%%%%%%%%%%%%%%%%%%%%%%%

%%%%%%%%%%%%%%%%%%%%%%%%% Tables %%%%%%%%%%%%%%%%%%%%%%%%%%%%
\begin{table}
\begin{tabular}{c|cccc|cccc|cccc}
 & \multicolumn{4}{c}{\hgo} & \multicolumn{4}{c}{\hgt} &
\multicolumn{4}{c}{\hgf} \\
 &
 K=0  & K=1  & K=2  & K=3 &
 K=0  & K=1  & K=2  & K=3 &
 K=0  & K=1  & K=2  & K=3 \\ \hline
E [ MeV ] &
1.37 & 1.45 & 1.20 & 1.52 &
1.55 & 1.58 & 1.18 & 1.53 &
1.83 & 1.62 & 1.14 & 1.53 \\
$B(E3)/B(E3)_{\rm s.p.}$ &
6.6  & 11.9 & 10.0 & 1.0  &
7.6  & 10.1 & 10.1 & 0.8  &
11.5 & 11.2 & 10.2 & 0.7  \\ \hline
\end{tabular}
\caption{
Calculated excitation energies of octupole vibrations and
$B(E3; 0^+ \rightarrow 3^-,K)$ values estimated
using the strong coupling scheme for SD $^{190,192,194}$Hg.
}
\label{oct-E-E3}
\end{table}

\mediumtext
\begin{table}
\begin{tabular}{c|ccc}
 & \hgo & \hgt & \hgf \\ \hline
E [ MeV ] & 1.39 & 1.50 & 1.45 \\
$B(E2)/B(E2)_{\rm s.p.}$ &
2.7 & 3.0 & 3.8 \\ \hline
\end{tabular}
\caption{
Calculated excitation energies of $\gamma$ vibrations and
$B(E2; 0^+ \rightarrow 2^+,K=2)$ values estimated
using the strong coupling scheme for SD $^{190,192,194}$Hg.
}
\label{gamma-E-E2}
\end{table}

\widetext\squeezetable
\begin{table}
\begin{tabular}{ll|ll|ll}
   &     & \multicolumn{2}{c}{$\pi=+$} & \multicolumn{2}{c}{$\pi=-$} \\
     &           & The lowest & The second & The lowest & The second \\
\hline
\hgo & $E_x'$ [ keV ] & 113        & 389        & $\approx 0$ & 256 \\
     & config.   & $\nu(7_1 \otimes 7_2)_{\alpha=0}$
                 & $\nu(7_1 \otimes [505\ 11/2])_{\alpha=0,1}$
                 & $(\mbox{oct.vib.})_{\alpha=1}$
                 & $\nu(7_1 \otimes [642\ 3/2])_{\alpha=0}$ \\
                  \cline{2-6}
     & exp.      & Band 3     &            & Band 2     & Band 4 \\
\hline
\hgt & $E_x'$ [ keV ] & 611        & 611        & 441        & 632 \\
     & config.   & $\nu(7_1 \otimes [512\ 5/2])_{\alpha=1}$
                 & $\nu(7_1 \otimes [512\ 5/2])_{\alpha=0}$
                 & $\nu(7_1 \otimes [642\ 3/2])_{\alpha=1}$
                 & $\nu(7_1 \otimes [642\ 3/2])_{\alpha=0}$ \\
                  \cline{2-6}
     & exp.      &            &            & Band 2     & Band 3 \\
\hline
\hgf & $E_x'$ [ keV ] & 857        & 892        & 738        & 759 \\
     & config.   & $\nu([514\ 7/2])^2_{\alpha=0}$
                 & $\pi([530\ 1/2])^2_{\alpha=0}$
                 & $(\mbox{oct.vib.})_{\alpha=0}$
                 & $(\mbox{oct.vib.})_{\alpha=1}$ \\
                  \cline{2-6}
     & exp.      &            &            & Band 2     & Band 3 \\
\hline
\end{tabular}
\caption{
The lowest and the second lowest configurations
at $\hbar\w=0.4\mbox{MeV}$ in each parity sector.
The proposed assignments of the observed excited SD bands are also shown.
The excitation energies of the negative-parity two-quasineutron states,
256keV for \hgo\ and 441 and 632keV for \hgt,
contain very weak octupole correlations.
The corresponding unperturbed two-quasineutron energies are
261, 460 and 635 keV, respectively.
}
\label{Ex}
\end{table}

%%%%%%%%%%%%%%% Figures %%%%%%%%%%%%%%%%%%%%%%%%%%%%%%%
\begin{figure}[tbp]
\caption{
Kinematic (solid lines) and dynamic (dashed lines) moments of inertia
for SD $^{152}$Dy
calculated in the cranked Nilsson Hamiltonian with (thick lines)
and without (thin lines)
the additional term $h_{\rm add}$ in eq.(\protect\ref{additional-term}).
The rigid-body and the Inglis moments of inertia are shown by
dash-dotted and dotted lines, respectively.
The parameters used in the calculation are the same as those used in
Ref.\protect\cite{Nak95} and pairing correlations are neglected.
Symbols are experimental $\J$ taken from Ref.\protect\cite{Dag95}.
}
\label{dy152_J}
\end{figure}

\begin{figure}[tbp]
\caption{
Quasiparticle routhians for neutrons (left)
and protons (right) in \hgo.
The top parts show the routhians in the Nilsson potential
without the additional term $h_{\rm add}$,
the middle for those with $h_{\rm add}$,
and the bottom for those in the Woods-Saxon potential with the
``universal'' parameters.
Solid, dashed, dotted and dash-dotted lines correspond to quasiparticles
with $(\pi,\alpha)=(+,-1/2)$, $(+,1/2)$, $(-,-1/2)$ and $(-,1/2)$,
respectively.
See text for details.
}
\label{hg190_routh_comp}
\end{figure}

\begin{figure}[tbp]
\caption{
Neutron quasiparticle routhians in the Nilsson potential with
$h_{\rm add}$ for SD $^{190,192,194}$Hg.
See text and caption to Fig.~\protect\ref{hg190_routh_comp}
for details.
}
\label{neutron_routhians}
\end{figure}

\begin{figure}[tbp]
\caption{
Calculated RPA eigen-energies of negative-parity states
for SD \hgo, plotted as functions of rotational frequency.
Open (filled) circles indicate states with signature $\alpha=0$
($\alpha=1$).
Large, medium, and small circles indicate RPA solutions with $E3$
transition amplitudes
$\left( \sum_K |\bra{n}Q_{3K}^e\ket{\w}|^2\right)^{1/2}$
larger than 200 efm$^3$, larger than 100 efm$^3$ and
less than 100 efm$^3$, respectively.
Note that routhians for the yrast SD band correspond to the horizontal
axis ($E'_x=0$).
The observed routhians for Band 2\protect\cite{Cro95}
are shown by open squares.
}
\label{oct_hg190}
\end{figure}

\begin{figure}[tbp]
\caption{
The same as Fig.~\protect\ref{oct_hg190}, but for \hgt.
}
\label{oct_hg192}
\end{figure}

\begin{figure}[tbp]
\caption{
The same as Fig.~\protect\ref{oct_hg190}, but for \hgf.
}
\label{oct_hg194}
\end{figure}

\begin{figure}[tbp]
\caption{
Calculated RPA eigen-energies for $\gamma$ vibrational states for SD
\hgo, plotted as functions of rotational frequency.
The lower part is for the signature $\alpha=0$ routhians and the upper
for the $\alpha=1$.
Large filled, small filled and small open circles indicate the $\gamma$
vibrational states whose $K=2$ $E2$ amplitudes
$|\bra{n}Q_{22}^e\ket{\w}|$
are larger than 20 efm$^2$, larger than 10 efm$^2$ and
less than 10 efm$^2$, respectively.
The unperturbed two-quasineutron (two-quasiproton) routhians are also shown
by solid (dashed) lines.
}
\label{quad_hg190}
\end{figure}

\begin{figure}[tbp]
\caption{
The same as Fig. \protect\ref{quad_hg190}, but for \hgt.
}
\label{quad_hg192}
\end{figure}

\begin{figure}[tbp]
\caption{
The same as Fig. \protect\ref{quad_hg190}, but for \hgf.
}
\label{quad_hg194}
\end{figure}

\begin{figure}[tbp]
\caption{
Calculated (solid lines) and experimental (symbols) dynamic moments of
inertia for excited SD bands in \hgo\ (left), \hgt\ (middle) and
\hgf\ (right).
$\J$ for the yrast SD bands are also displayed at the top.
Dotted lines indicate the yrast $\J$, which are approximated by
the Harris formula (\protect\ref{Harris}).
The parameters, $J_0$, $J_1$ and $J_2$ used in the formula
are shown in units of $\hbar^2\mbox{MeV}^{-1}$,
$\hbar^4\mbox{MeV}^{-3}$ and $\hbar^6\mbox{MeV}^{-5}$, respectively.
}
\label{J2}
\end{figure}

\begin{figure}[tbp]
\caption{
Calculated (solid lines) and experimental (symbols) dynamic moments of
inertia for excited SD bands in \hgf.
Thin solid lines are the same as in Fig.~\protect\ref{J2},
while thick lines indicate the results obtained by using the
slightly stronger coupling strengths ($f_3=1.05$)
for the octupole interactions.
Dotted lines indicate the $\J$ for the yrast SD band (see caption to
Fig. \protect\ref{J2}).
}
\label{J2_hg194}
\end{figure}

\begin{figure}[tbp]
\caption{
Electric $E3$ transition amplitudes,
$\left| t[\frac{1}{2}(1+\tau_3)Q_{3K}^\alpha]\right|
 = |\bra{\w}Q_{3K}^e \ket{n}|$,
for the lowest RPA solutions with the signature $\alpha=0$ (lower)
and the $\alpha=1$ (upper) for \hgo\ (left), \hgt\ (middle) and
\hgf\ (right).
$K=0$, 1, 2 and 3 components are denoted by solid, dashed, dotted and
dash-dotted lines, respectively.
Total values (thick solid lines) are defined by
$\left( \sum_K |\bra{\w}Q_{3K}^e \ket{n}|^2 \right)^{1/2}$.
}
\label{E3_amplitudes}
\end{figure}

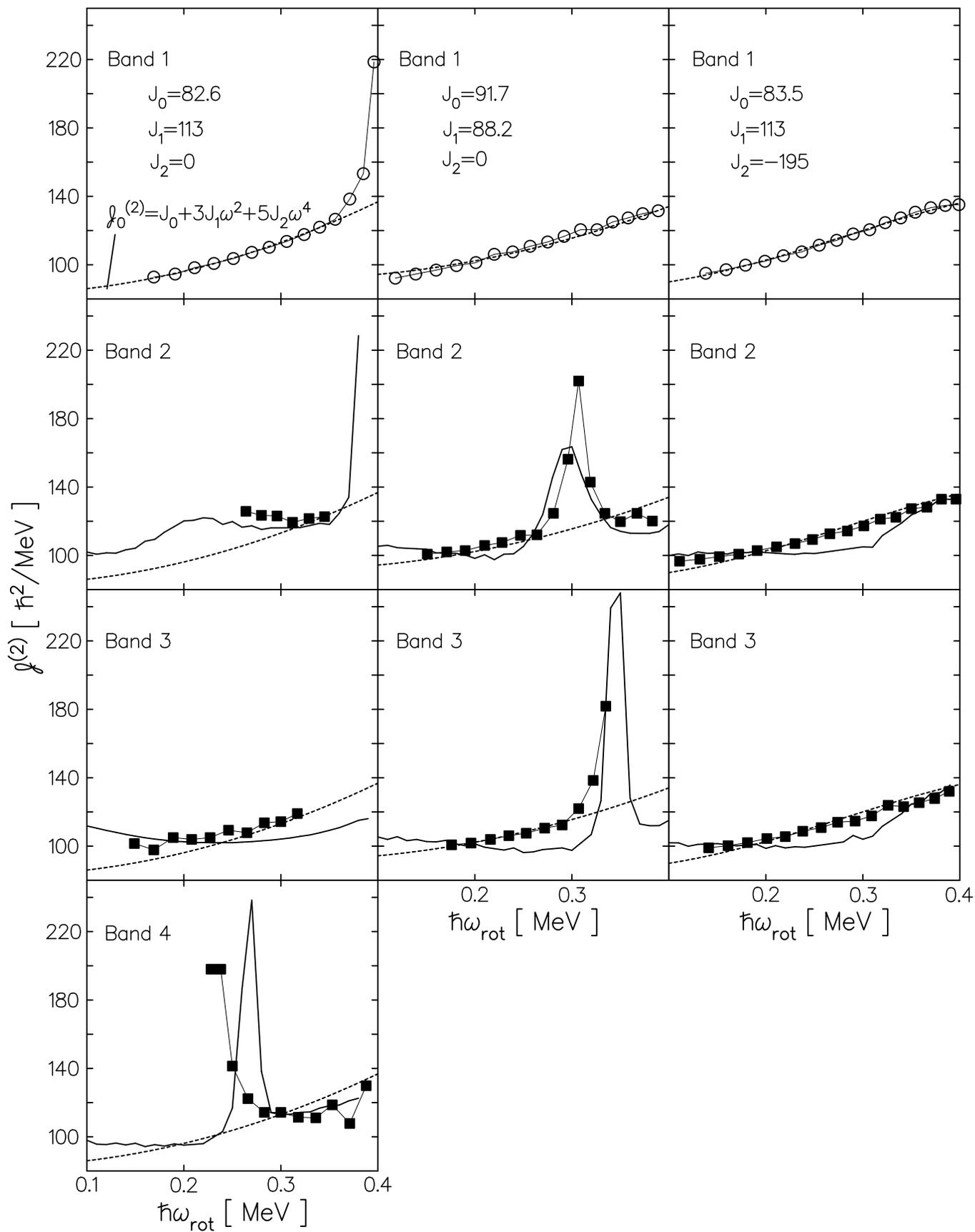
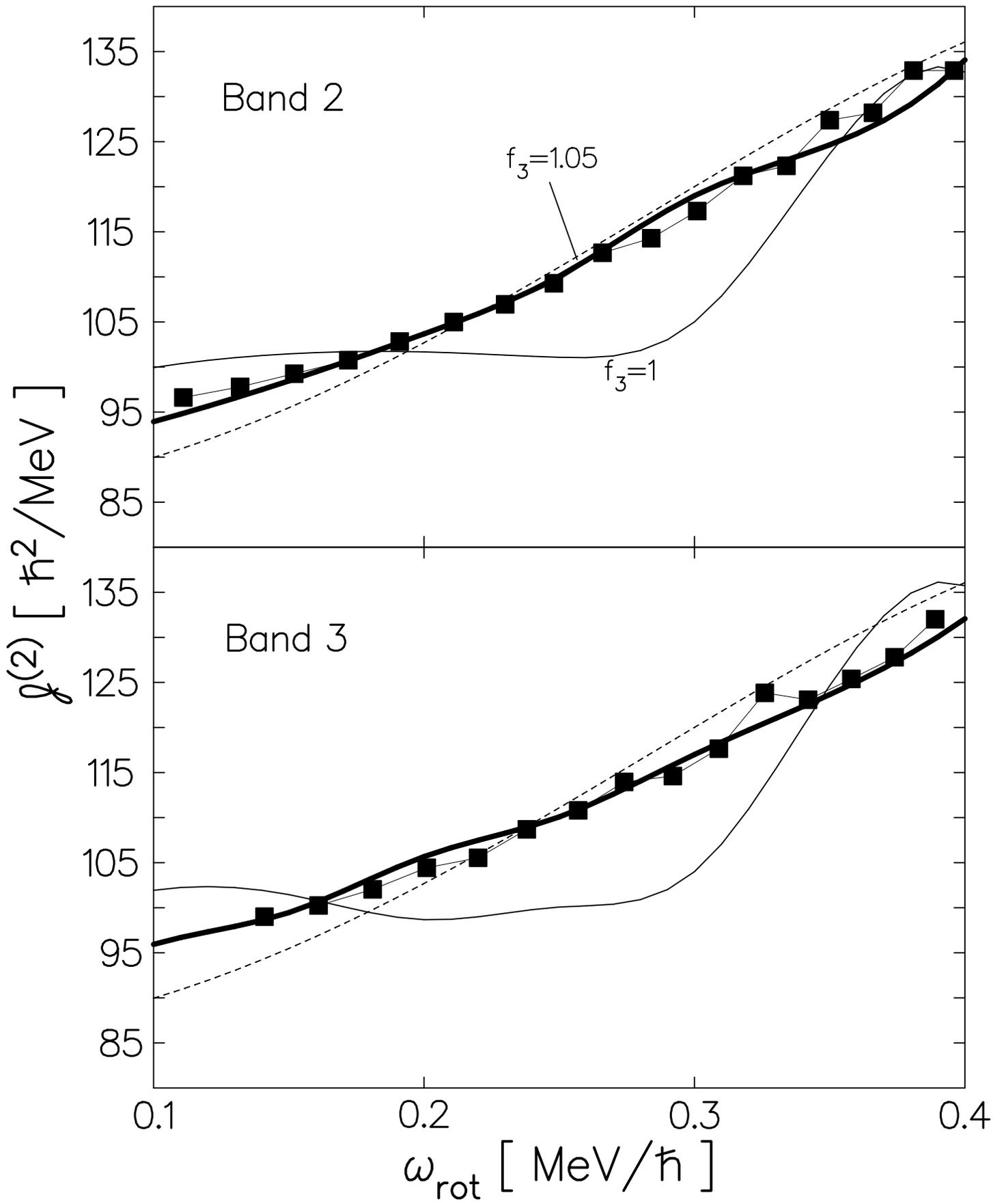
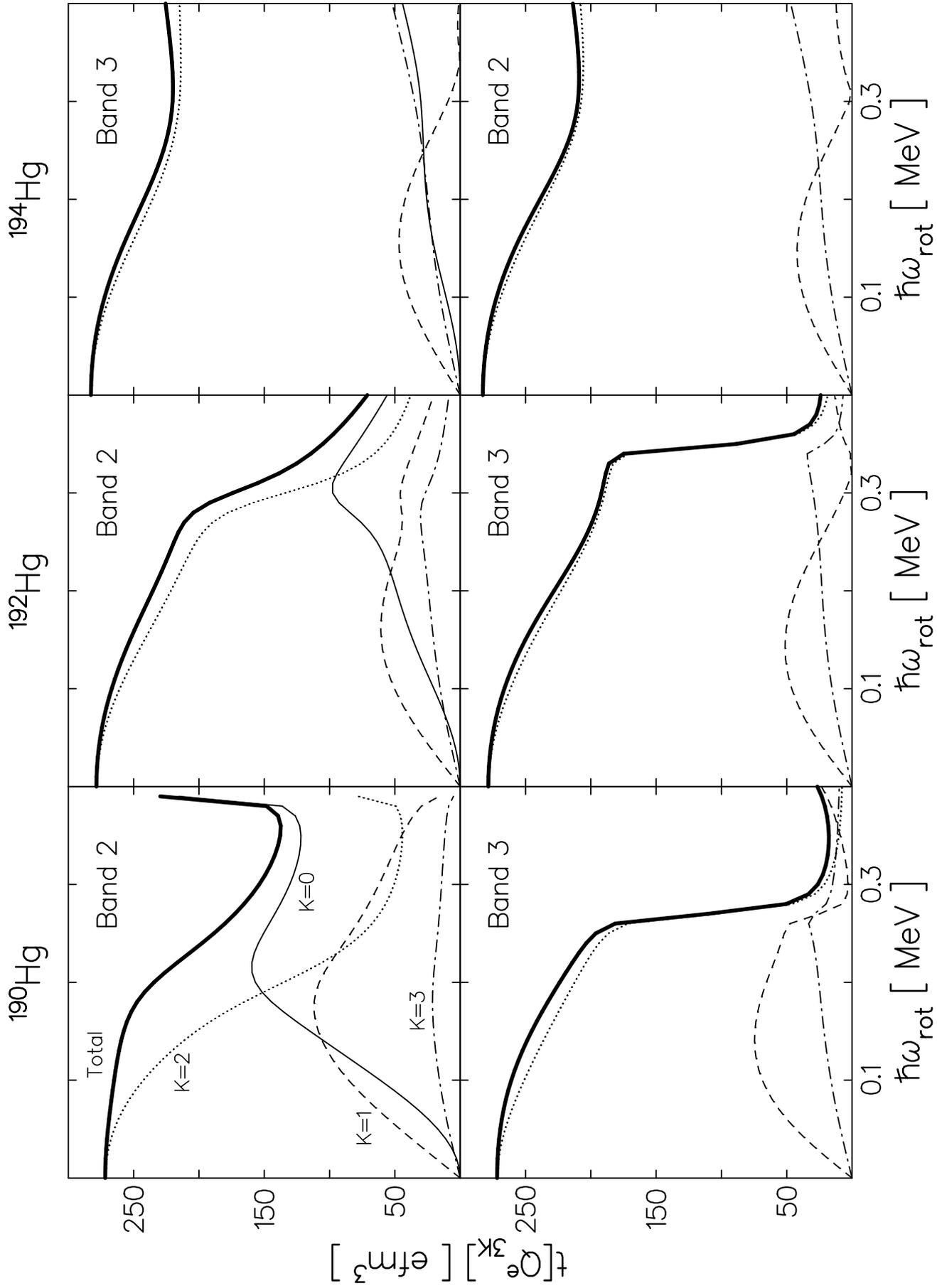
\end{document}